\documentclass[amsmath,amssymb,floatfix,prl,epsf,twocolumn,superscriptaddress]{revtex4}
\usepackage{amsmath,amssymb,natbib,bm,graphicx,url,epsfig,wasysym}
\usepackage[ansinew]{inputenc}
\usepackage{bm}
\usepackage{color}
\newcommand{\be}{\begin{equation}}
\newcommand{\ee}{\end{equation}}
\newcommand{\bes}{\begin{equation}\begin{split}}
\newcommand{\ees}{\end{split}\end{equation}}
\newcommand{\bea}{\begin{eqnarray}}
\newcommand{\eea}{\end{eqnarray}}
\newcommand{\nn}{\nonumber}

\DeclareMathOperator{\curl}{curl}

\def\beq{\begin{equation}}
\def\eeq{\end{equation}}
\def\bea{\begin{eqnarray}}
\def\eea{\end{eqnarray}}

\begin{document}


\title{ Spontaneous formation of a non-uniform chiral spin liquid in  moat-band lattices
}

\author{Tigran A. Sedrakyan}

\affiliation{William I. Fine Theoretical Physics Institute and Department of Physics,  University of Minnesota, Minneapolis, Minnesota 55455, USA}
\affiliation{Physics Frontier Center and Joint Quantum Institute, University of Maryland, College Park, Maryland 20742, USA}

\author{Leonid I. Glazman}

\affiliation{Department of Physics, Yale University, New Haven, Connecticut 06520, USA}

\author{Alex Kamenev}

\affiliation{William I. Fine Theoretical Physics Institute and Department of Physics,  University of Minnesota, Minneapolis, Minnesota 55455, USA}

\begin{abstract}
 A number of lattices exhibit moat-like band structures, i.e. a band with infinitely degenerate energy minima attained along a closed  line in the Brillouin zone. If such a  lattice is populated with hard-core bosons, the degeneracy prevents their condensation. At half-filling, the system is equivalent to $s=1/2$ XY model at zero magnetic field, while absence of condensation translates into the absence of magnetic order in the XY plane. Here we show that the ground state  breaks the time-reversal as well as inversion symmetries. This state, which may be identified with the  chiral spin liquid, has a bulk gap and chiral gapless edge excitations.   The applications of the developed analytical theory include an explanation of recent numerical findings and a suggestion for the chiral spin liquid realizations in experiments with cold atoms in optical lattices. 
\end{abstract}

\date{\today}

\maketitle

Recent experiments with Herbertsmithite and other zinc paratacamites~\cite{Helton-2007,Vries-2008,Helton-2010,Han-2012,pilon} indicated a possible observation of a long-sought quantum spin liquid. 
These developments added motivation for theoretical investigation of frustrated spin Hamiltonians on a variety of lattices, see e.g. \cite{Varney-2012,  Motrunich-2-2013, Motrunich-2014, Huse, Balents, Sachdev-2}, 
including triangular, honeycomb, and Kagome ones. In a majority of models, an increase of the frustration could lead to a formation of $Z_2$ quantum spin liquid preserving time-reversal symmetry (TRS). In recent numerical studies~\cite{CSL-kagome, kagome-heis,Ludwig-2014} of a frustrated Kagome lattice, however, a chiral spin liquid ground state was found.

Chiral spin liquid (CSL) suggested by Kalmeyer and Laughlin~\cite{Kalmeyer-1987}, is 
a possible ground state of frustrated spin systems, resembling 
quantum Hall states.   
CSL is gapped in the bulk, but supports gapless chiral excitations along edges, owing to a nontrivial Chern number.  Such a ground state is described by an incompressible bosonic wave function\cite{Kalmeyer-1987}.
The concept was further developed  in Ref.~[\onlinecite{WWZ}], where a set of order parameters was identified and an explicit solvable example of a CSL was constructed. A convenient choice for a CSL  order parameter is the chirality, $\chi= \langle{\bf S}_i \cdot ({\bf S}_j \times {\bf S}_k)\rangle $, 
where ${\bf S}$ is the spin-1/2 operator and $i,j,k$ label lattice cites forming a triangular plaquet. 
Finite expectation value, $\chi$, violates P and T symmetries, leaving the combined PT symmetry intact.

The crucial observation is that once the frustration is strong enough, the lattice dispersion exhibits the {\em moat} shape - i.e. the degenerate energy minimum along a closed line in the Brillouin zone \cite{we2}. In such a case, as in one dimension (1D), the single particle density of states diverges at the bottom of the band. In analogy with the 1D Tonks-Girardeau gas, it suggests to transform the bosons into {\em spinless} fermions\cite{we2,we},  which automatically satisfy the hard core condition. In two dimensions it may be achieved with the help of the  Chern-Simons (CS) flux attachment, similar to the one employed in the theory of the fractional Quantum Hall effect \cite{Jain, lopez, Halperin}. Here we show that the CS transformation {\em on a moat-lattice}  
leads to fermions subject to fluxes staggered within the unit cell.  
Such staggered fluxes bring about topologically non-trivial fermionic ground state with a non-zero Chern number and chiral edge  states \cite{Haldane}.  
This is the main result of the present letter, which paves the way to most direct and intuitive identification 
of  CSL in magnetic systems as well as its realizations in cold atomic setup.

As an example we consider an antiferromagnetic (AF) $XY$ model on a honeycomb lattice with nearest-neighbor (NN) spin 
coupling, $J_1$, and next-NN coupling, $J_2$. Finite $J_2$ leads to frustration and hence the model  is expected to have a rich phase diagram. 
Numerical studies, performed using exact diagonalization~\cite{Varney-2012},
variational Monte Carlo\cite{carasquilla}, and the density matrix renormalization group (DMRG)~\cite{ZHW} techniques suggest that the AF order in X-Y plane survives up to a critical value $J_2/J_1\simeq 0.2$, where the system undergoes a phase transition. Reference~[\onlinecite{Varney-2012}] reported exact diagonalization study, suggesting existence of a new phase in the parameter range $ 0.2  \lesssim  J_2/J_1 \lesssim 0.36$. The authors suggested that it is a "Bose liquid" phase in which the spin ordering  is  absent down to  zero temperature. 
Recently Zhu, Huse,  and  White (ZHW) ~\cite{ZHW} employed DMRG technique in cylindrical geometry for the same 
range of parameters (throughout this paper we refer to it as intermediate frustration regime). They found that while 
there is indeed no order vis-a-vis the X-Y plane, there is a weak antiferromagnetic Ising order in $z$-direction. It breaks the symmetry between A and B sublattices  with $\langle S^z_A-S^z_B\rangle$ taking values in between $0.27$ and $0.28$. 

\begin{figure}[t]
\centerline{\includegraphics[width=90mm,angle=0,clip]{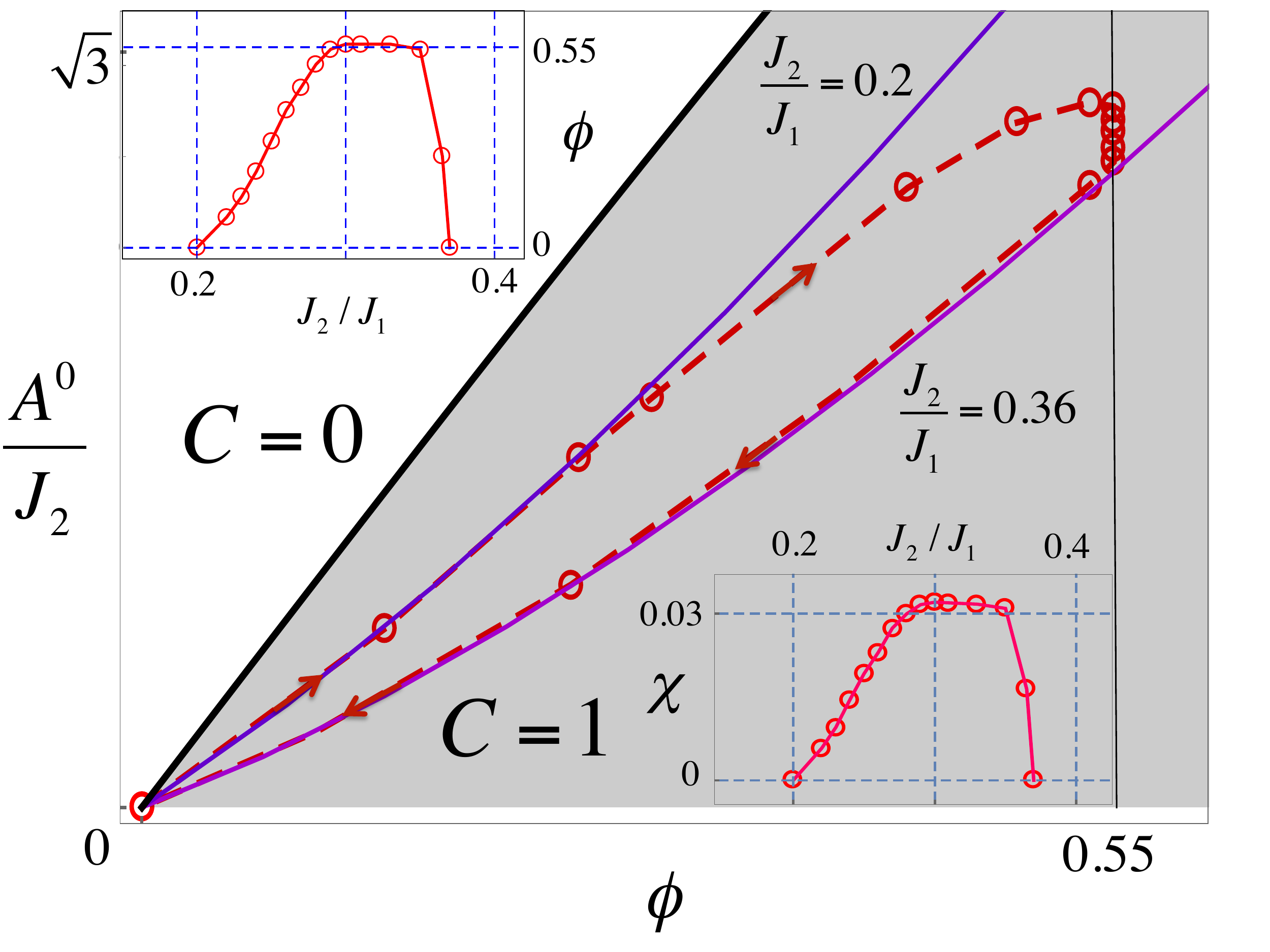}}
\caption{(Color online) Phase diagram in the space of dimensionless parameters $( A^0/J_2;\phi)$. The shadowed region, bounded by the bold line $A^0/J_2=3 \sqrt{3} \sin\phi$, represents topologically nontrivial states with Chern number $C=1$ \cite{Haldane}. The solutions of the self-consistency conditions ({\ref{EM}}), shown by the dashed line, all fall into the topological sector with $C=1$. Such solutions exist  
in the interval $0.2< J_2/J_1<0.36$, and are confined between the two  full thin lines representing  Eq.~(\ref{tildnu0}) for the boundaries of this interval $J_2/J_1=0.2; 0.36$.   Upper (lower) inset: staggered flux, $\phi$, (chirality, $\chi$,) vs. $J_2/J_1$,  as obtained from Eqs.~(\ref{EM}).}
\label{fig-gap}
\end{figure}

Here we study the model by  reformulating it as  a CS fermionic field theory on a lattice. We found that 
in the intermediate frustration regime the AF Ising order of ZHW is stabilized by appearence of staggered CS fluxes within the unit cell.  The zero-average modulated fluxes, induced by the lattice CS field, are exactly the same as postulated in a celebrated Haldane  model \cite{Haldane}. Solution of self-consistent mean-field equations puts  the model into its topologically non-trivial sector with Chern number $C=\pm 1$. This allows us to identify the $z$-modulated state of ZHW with CSL state, which supports gapless spinon excitations along the edges. It would be 
extremely interesting to see if DMRG studies can check this prediction.  Moreover we predict a relation between the AF Ising order parameter $ \phi=\frac{2\pi}{3} \langle S^z_A-S^z_B\rangle$ and chirality $\chi=\langle {\bf S}_{A} \cdot ({\bf S}_{B} \times {\bf S}_{A^\prime})  \rangle \propto \sin\phi$, reflected in instets of Fig.~\ref{fig-gap}, which may be also directly checked in simulations.

\begin{figure}[t]
\centerline{\includegraphics[width=70mm,angle=0,clip]{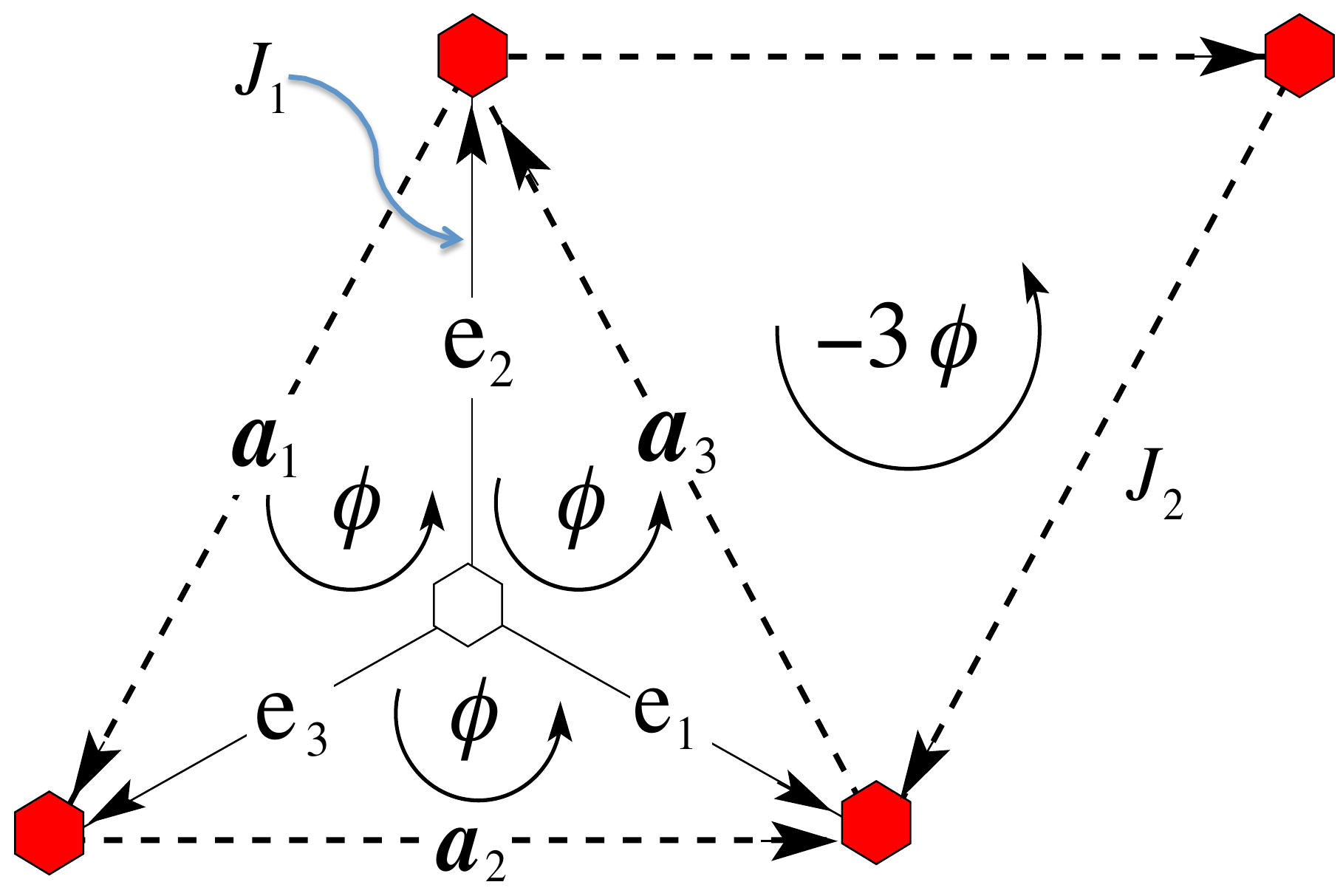}}
\caption{(Color online) Unit cell of honeycomb lattice with NN  $J_1$ and next NN $J_2$ couplings. The CS fluxes associated with spontaneously broken time-reversal symmetry are shown.  
}
\label{fig-phases-2}
\end{figure}

To quantify the aforementioned ideas, we start with the Hamiltonian of frustrated spin-1/2 XY model  on a honeycomb lattice with 
nearest and  next to nearest
neighbor interaction terms
\be
\label{H-XY}
H= J_1 \sum_{{\bf r},j}S^{+}_{\bf r}S^{-}_{{\bf r}+{\bf e}_j}+
J_2 \sum_{{\bf r},j}S^{+}_{\bf r}S^{-}_{{\bf r}+{\bf a}_j}+   H.c. 
\ee
Here the spin-1/2 operators are related to Pauli matrices as $S^{\pm}_{\bf r}\equiv \sigma_{\bf r}^{+}$ and $S^z_{\bf r}\equiv\sigma^{z}_{\bf r}/2$. 
Vectors ${\bf e}_j$ and ${\bf a}_j,\; j=1,2,3$, shown in Fig.~\ref{fig-phases-2}, are connecting  nearest 
and next-to-nearest neighbor cites of the honeycomb lattice.


The XY model (\ref{H-XY}) is equivalent to the model of hard-core bosons, as one may rewrite the spin 1/2 operators $S^{\pm}_{\bf r}$ in terms of bosonic creation and annihilation operators.  When $J_2/J_1>1/6$, the corresponding single particle dispersion relation 
undergoes dramatic changes: it becomes infinitively degenerate and exhibits an energy minimum along a closed line in the reciprocal space surrounding the $\Gamma$ point~\cite{we2} -- the {\em moat}. The single particle density of states diverges near the moat bottom as $(E-E_c)^{-1/2},$ highlighting similarities with 1D systems, where the ground state of hard-core bosons is given by the Tonks-Girardeau gas of free fermions. This observation supports the idea that {\em spineless fermion} representation might be an effective description of 2D boson systems in a moat, as was suggested in Refs.~[\onlinecite{we,we2}].  Advantage of spineless fermions is that they automatically satisfy the  hard-core condition and thus do not suffer from a repulsive interaction energy. 

We proceed with the lattice version of the CS transformation (its continium analog was emloyed in e.g. Refs.~[\onlinecite{Jain, lopez, Halperin,shan,read}])
\bea
\label{CS}
S_{\bf r}^{(\pm)}= c_{\bf r}^{(\pm)}\, e^{\pm i \sum_{{\bf r'}\neq {\bf r}} \arg[{\bf r}-{\bf r^\prime}] n_{\bf r'}}\,,
\eea
where the summation runs over all sites of the lattice. Since the bosonic operators on different sites commute,
the newly defined operators $c_{\bf r}$ and $c_{\bf r}^{\dagger}$  obey fermionic commutation relations.
Also notice that the number operator is given by $n_{\bf r} = c_{\bf r}^\dagger c_{\bf r}=S^z_{\bf r}+1/2$. 


Substitution of transformation Eq.~(\ref{CS}) into the Hamiltonian (\ref{H-XY}) yields

\bea
\label{H2}
H\!=\!  J_1 \sum_{{\bf r},j}c^\dagger_{\bf r}c_{{\bf r}+{\bf e}_j}e^{i {\cal A}_{ {\bf r}, {\bf r}+{\bf e_j}} }
\!+\!J_2 \sum_{{\bf r},j}c^\dagger_{\bf r} c_{{\bf r}+{\bf a}_j} e^{i{\cal A}_{ {\bf r}, {\bf r}+{\bf a_j}}}\!+\!H.c. \nn
\eea
where ${\cal A}_{{{\bf r}_1}, {\bf r}_2}=\sum_{r}\left[\arg({\bf r}_1-{\bf r})-\arg({\bf r}_2-{\bf r})\right] n_{\bf r}
$ with summation running over all lattice sites.  Due to  (spinless) fermionic nature of the operators the hard core condition is  automatically satisfied. 
One can remove exponential string operators by introducing CS magnetic field, ${\cal B}_{\bf r}={\cal A}_{{\bf r}+{\bf e}_1,{\bf r}+{{\bf e}_2}}+{\cal A}_{{\bf r}+{\bf e}_2,{\bf r}+{{\bf e}_3}}+{\cal A}_{{\bf r}+{\bf e}_3,{\bf r}+{{\bf e}_1}}=2\pi n_{\bf r}$, which is the lattice analog of  ${\cal B}_{\bf r}=\curl {\cal A}$ (see Fig.~\ref{fig-phases-2}). To this end one introduces the 
 $\delta$-function, $2\pi \delta({\cal B}_{\bf r}/(2\pi)- n_{\bf r})=\int\prod_{\bf r}[dA^0_{\bf r}]\exp\left[iA^0_{\bf r}({\cal B}_{\bf r}/(2\pi)- n_{\bf r})\right]$.
The corresponding functional integration with respect to the 
CS vector potential ${\cal A}_{{\bf r}_1,{\bf r}_2}$ is also implied. 
The Lagrange multiplier $A^0_{\bf r}$ plays the role of the zero component of the vector potential.

These notations enable one to represent the model
as a fermion system coupled to the fluctuating CS gauge field.  
In analogy with the continuum case~[\onlinecite{Halperin}],  we write
\bea
\label{H21}
&&S= \int dt\Biggl[\sum_{{\bf r}}\bar {c}_{\bf r}\left(i\partial_t-A^0_{\bf r}\right)c_{\bf r}+\frac{1}{2\pi} \sum_{{\bf r}} A^0_{\bf r}{\cal B}_{\bf r}\\
&&\!-J_1 \sum_{{\bf r},j}{\bar c}_{\bf r}c_{{\bf r}\!+\!{\bf e}_j}e^{i {\cal A}_{ {\bf r}, {\bf r}+{\bf e_j} }}
\!-\! J_2 \sum_{{\bf r},j}{\bar c}_{\bf r} c_{{\bf r}\!+\!{\bf a}_j} e^{i {\cal A}_{ {\bf r}, {\bf r}+{\bf a_j} }}
+H.c.\!\Biggr]\nn.
\eea
Here the fermions are Gaussian and one may integrate them out obtaining the fermionic free energy functional, $W[ A^0_{\bf r},{\cal B}_{\bf r}]$. Along with the CS term the latter  defines the formally exact 
effective action,
$S_{eff}=W[A^0_{\bf r},{\cal B}_{\bf r}]+\frac{1}{2\pi} \int dt \sum_{{\bf r}} A^0_{\bf r}{\cal B}_{\bf r}$.
We shall treat it in the mean-field approximation, looking for solutions of the equations of motion:  $\delta_{A^0_{\bf r}}S_{eff}=0$, $\delta_{{\cal B}_{\bf r}}S_{eff}=0$.
Non-trivial solutions of these equations, if exist, determine expectation values of CS fields ${\cal B}_{\bf r} $ and $A^0_{\bf r}$ in a self-consistent way. While the mean-field is uncontrolled,  it will result in a  solution with the fermionic  spectrum gapped in the bulk. The fluctuation corrections are thus finite and appear to be numerically small, giving some confidence in the validity of the mean-field. In this sense situation here is more favorable as compared to the theory of the half filled Landau level\cite{Jain,Halperin}, where the spectrum is gapless. 

We shall look for  spatially homogeneous  solutions of the mean-field equations, allowing for broken symmetry between  
$A$ and $B$ sub-lattices. Such choice is motivated by recent numerical results of ZHW where an
asymmetry between sublattices $A$ and $B$ was observed. Therefore, we will look for solution 
$ \left\{A_{{\bf r}_A}^0,{\cal B}_{{\bf r}_A}\right\}$ and $ \left\{A_{{\bf r}_B}^0,{\cal B}_{{\bf r}_B}\right\}$
that are independent of ${\bf r}_A$ and ${\bf r}_B$ respectively. It is convenient to separate 
symmetric and antisymmetric combinations of gauge fields
between   $A$ and $B$ sites, belonging to the same unit cell:
$(A_{{\bf r}_A}^0 \pm A_{{\bf r}_B}^0) $
and $({\cal B}_{{\bf r}_A} \pm {\cal B}_{{\bf r}_B}) $.  
It is easy to see that the homogeneous symmetric components 
may be gauged out at half filling\cite{we2}. The reason is that the corresponding total flux threading the unit cell  is $2\pi$, which may be disregarded due to the periodicity. 
This leads to the conclusion that the total CS flux threading the unit cell is gauge equivalent to zero.  Hence one can gauge out the
phases ${\cal A}_{ {\bf r}, {\bf r}+{\bf e_j} }$ residing on NN links in Eq.~(\ref{H21}). 
As a result only antisymmetric gauge fields residing on the next-NN links (see Fig.~\ref{fig-phases-2}) remain. These fields 
give rise to the flux threading the large equilateral triangle with a site in it, depicted in Fig.~\ref{fig-phases-2}.
It is given in terms of antisymmetric fields  as ${3\phi}\equiv({\cal B}_{{\bf r}_A} - {\cal B}_{{\bf r}_B})$. The flux threading 
the neighboring "empty" equilateral triangle (with no site in it), is thus $-3\phi$, which is consistent with the fact that the total flux through the unit cell is zero. The antisymmetric component of the gauge field thus precisely gives raise to the staggered Haldane flux configuration\cite{Haldane} in the unit cell.

Introducing notation, $A^0=(A^0_{{\bf r}_A} - A^0_{{\bf r}_B})/2$, the equations of motion for the asymmetric components take the form
\bea
\label{EM}
\!\!\partial_{A^0}W(A^0,\phi)+\frac{3}{2\pi}\phi=0,\;\;\partial_{\phi}W(A^0,\phi)+\frac{3}{2\pi}A^0=0.
\eea
The field $A^0$ here plays the role of the inversion symmetry breaking mass term. Indeed, it 
appears in the action (\ref{H21}) as 
${A}^0\sum_{[{\bf r}_A,{\bf r}_B]}\left[\frac{3}{2\pi}{\phi} - \left(n_{{\bf r}_A}- n_{{\bf r}_B}\right)\right],$ where the summation is performed over NN dimer pairs $[{\bf r}_A,{\bf r}_B]$.  
Then, the first equation of motion (\ref{EM}) 
yields the self-consistency condition ${3\phi}=2\pi \langle n_{{\bf r}_A}-n_{{\bf r}_B}\rangle$.

The above consideration shows that  the fermionic mean-field theory with the staggered CS flux configuration depicted in Fig.~\ref{fig-phases-2}, naturally gives raise to the
Haldane's model\cite{Haldane},
studied in connection with parity anomaly in a honeycomb lattice.
The spectrum of the Haldane Hamiltonian is gapped, and consist of two bands: 
\bea
\label{bands}
 E_{\bf p}(A^0, \phi)=
  J_2 \cos\phi\left[G_{\bf p}^2-3\right]
  \pm\sqrt{m_{\bf p}^2+ J_1^2G_{\bf p}^2},
  \eea 
 where $ m_{\bf p}= A^0- 2 J_2\sin\phi \sum_j \sin({\bf p}{\bf a}_j)$
is the gap, and $G_{\bf p}=|\sum_{i=1}^3 e^{i {\bf p} {\bf e}_i}|$. 
\begin{figure}[t]
\centerline{\includegraphics[width=75mm,angle=0,clip]{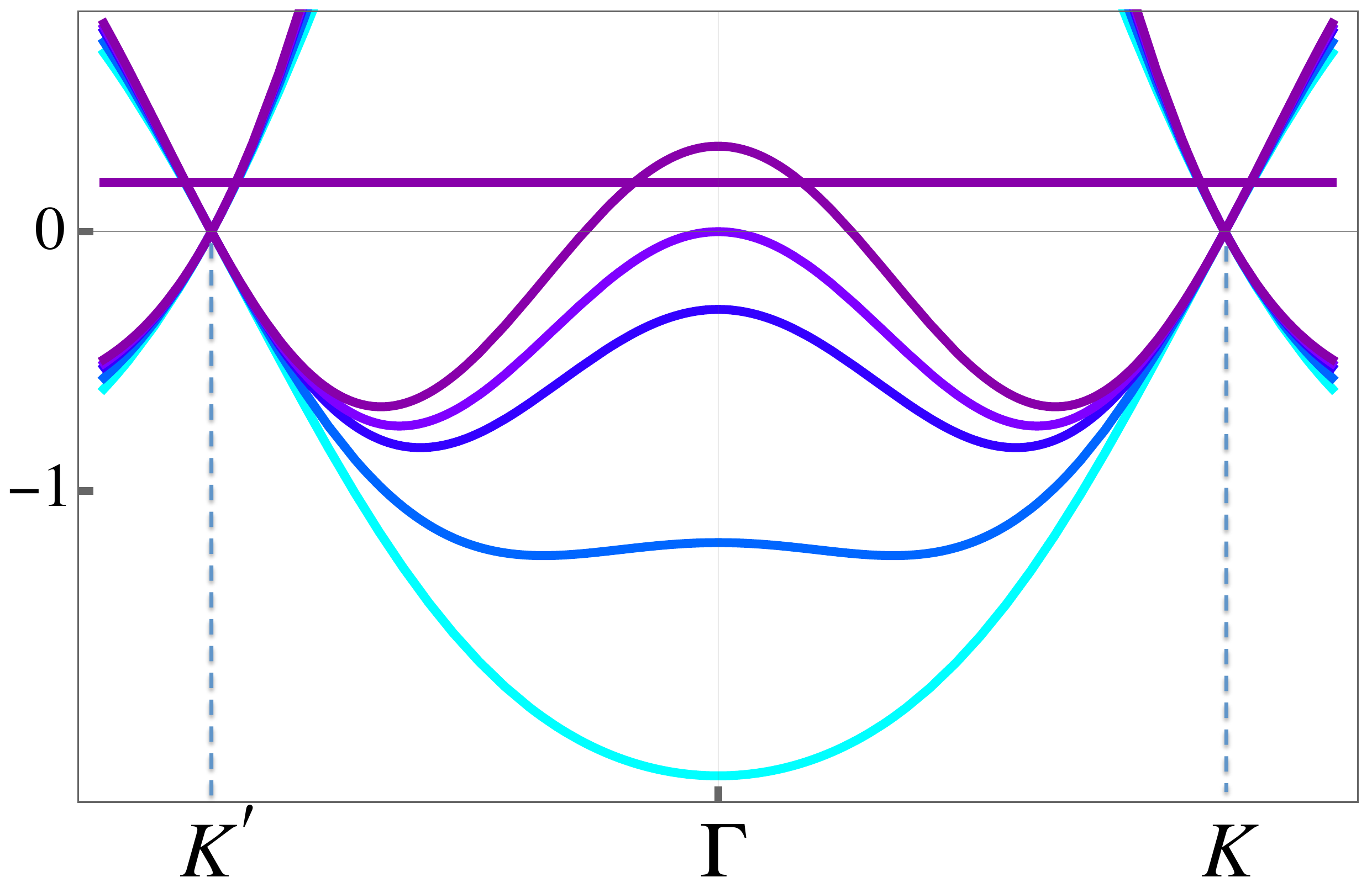}}
\caption{(Color online) Bare band spectrum $E_{{\bf p}}(0, 0)$ plotted from Eq.~(\ref{bands}) for  
$J_2/J_1=0.1; 0.2; 0.3; 1/3; 0.37 $ from bottom up. 
Horizontal line represents chemical potentials at half filling for $J_2/J_1=0.37$.  For $J_2/J_1\leq 1/3$, the chemical potential is at $\mu=0$.
For $J_2/J_1=1/3$ the maximum of the lowest branch at $\Gamma$ point reaches the hight of Dirac points $K$ and $K^\prime$.  A phase transition is thus expected for $J_2/J_1 \gtrsim1/3$,  towards the spin density wave state \cite{chubukov}.}
\label{fig-1}
\end{figure}
Fermionic ground state energy of the model  is given by  $W(A^0,\phi)=\sum_{\bf p} E_{\bf p}(A^0, \phi) $, where the sum runs over occupied states of the half-filled system.
For $J_2/J_1\leq1/3$, only the lowest band is filled, Fig.~\ref{fig-1}. 
Substituting $W(A^0,\phi)$, along  with the spectrum (\ref{bands}), into first Eq.~(\ref{EM}), one obtains a self 
consistent equation relating parameters 
$\phi$ and $A^0$:
\begin{eqnarray}
\label{tildnu0}
\frac{3}{2\pi}{\phi}=\sum_{\bf p} \frac{m_{\bf p}}{ \sqrt{m_{\bf p}^2+ J_1^2G_{\bf p}^2}}.
\end{eqnarray}
Numerical solution of Eq.~(\ref{tildnu0}) is shown in the Haldane's phase diagram in Fig.~\ref{fig-gap} 
for two values $J_2/J_1=0.2;0.36$.

The second of Eqs.~(\ref{EM})
minimizes the ground state energy $F =W( A^0, {\phi}) + \frac{3}{2\pi} \phi A^0 $ with respect to $\phi$ given the self-consistency relation (\ref{tildnu0}). Minimization yields 
finite values,  shown by dashed line in Fig.~\ref{fig-gap}, for both, $\phi$ and $A^0$ in the intermediate frustration regime $0.2\lesssim J_2/J_1\lesssim 0.36$. At the boundaries of this regime $\phi$ and $A^0$ vanish simultaneously, indicating two distinct phase transitions.   
In between these two transitions there is a broad  regime where $\phi\approx 0.55$ is almost independent of the ratio $J_2/J_1$. 
All these values fall in the topologically nontrivial region of the phase diagram depicted in Fig.~\ref{fig-gap}  
resulting thus in topological phase of fermions with Chern number $C=\pm1$. On the other hand, finite $\phi$ means stabilization of Ising antiferromagnetic ordering of the spins for $0.2\lesssim J_2/J_1\lesssim 0.36$, in agreement with the DMRG results of ZHW~\cite{ZHW}. 
Moreover, the state we suggest bears the hallmarks of a CSL. It is characterized by a finite gap for the $S=1$ excitations in the bulk and gapless chiral edge state having $S=1/2$ spinon excitations.

\begin{figure}[t]
\centerline{\includegraphics[width=85mm,angle=0,clip]{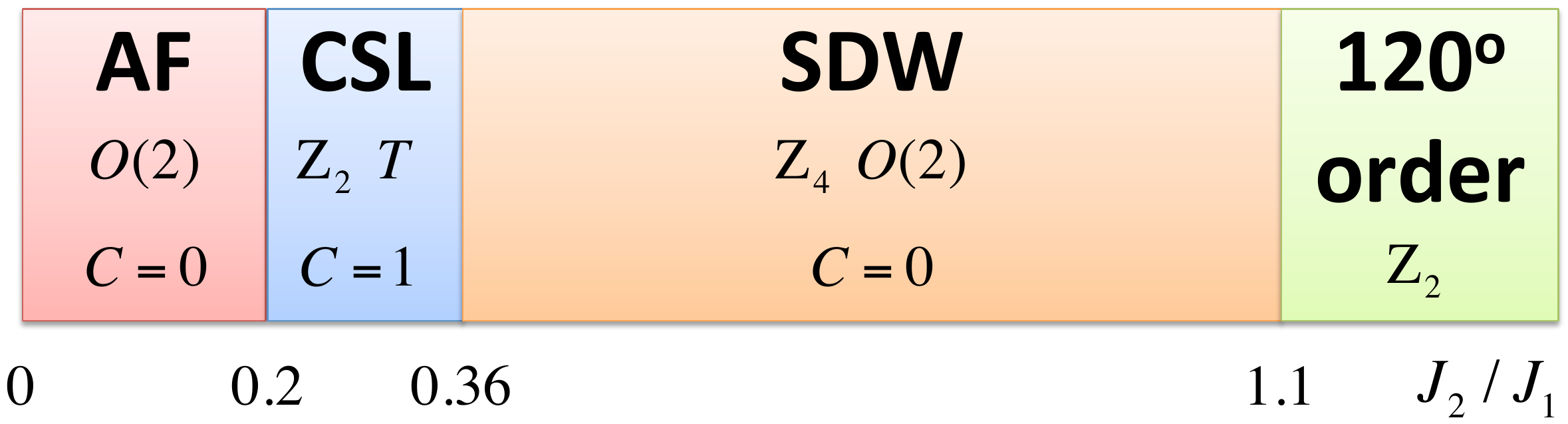}}
\caption{(Color online) Schematic phase diagram of the $XY$ model across the range of the ratio $J_2/J_1$.  Phases are marked by broken symmetries and corresponding Chern numbers, $C$.}
\label{PHASE1.pdf}
\end{figure}

The transition at $J_2/J_1\approx 0.2$ is continuous, while the one at $J_2/J_1\approx 0.36$ appears to be of the first order.  The latter is associated with the change of the underlying band structure. Indeed, at $J_2/J_1=1/3$ the maximum at $\Gamma$ point reaches the energy of the Dirac points at $K$ and $K'$, Fig.~\ref{fig-1}. Thus at  $J_2/J_1> 1/3$ the half-filled system exhibits hole fermi pocket at $\Gamma$ and particle fermi pockets at $K$ and $K'$. Presence of the occupied states in the upper band near  $K$ and $K'$ suppresses
the gap (since the upper band moves up in energy, the total energy is not  lowered). In our mean-field treatment the non-trivial self-consistent solution disappears (see insets to Fig.~\ref{fig-gap}) in the {\em first order} way at $J_2/J_1\approx 0.36$, giving rise to the phase diagram depicted in Fig.~\ref{PHASE1.pdf}\cite{footnote}.

The mean-field theory treatment of the intermediate frustration regime is in a rather satisfactory agreement with the numerical data. 
In the bulk of this range $\phi \approx 0.55 $ (see upper inset in Fig.~\ref{fig-gap}) which corresponds $  \langle S^z_{A}-S^z_{B}\rangle \approx 0.26$. This agrees (within small error bars) with the numerical observation reported in Ref.~[\onlinecite{ZHW}]. 
The numerical value of the ground state energy  per spin for e.g.  $J_2/J_1=0.3$ is $F/J_1\approx -0.311$.   
This is lower  than the corresponding exact diagonalization result~[\onlinecite{Varney-2012}] by only about $5\%$. 

In a finite-length cylinder geometry, spin transfer between the two edge states  reveals the fractional nature of the spinon excitations. In a numerical simulation, this can be checked by realizing Laughlin's adiabatic flux insertion procedure. 
The flux threading the cylinder  affects spins on $A$, $B$ sub-lattices differently.
This is the so called topological pump charge effect\cite{thou}, which gives possibility to directly measure the Chern number\cite{sheng,CSL-kagome,troyer}.  The latter can be obtained by applying $2\pi$ flux and calculating the polarization. 
The finite order parameter $\phi$  implies time reversal symmetry broken chiral ground state,
leading to a finite value of  chirality $\chi=\langle{\bf S}_A \cdot ({\bf S}_B \times {\bf S}_{A^\prime}\rangle)$, where the spins reside on the sites of the $120^\circ$ triangle shown in Fig.~\ref{fig-phases-2}. 
The result is plotted in the lower inset in  Fig.~\ref{fig-gap}, showing that one is expected to find $\chi\approx \pm 0.03$ in the bulk of the CSL range.  This may be also directly checked by numerics. 
The transition  into CSL at  $J_2/J_1\approx 0.2$ described in terms of two order parameters $A^0$  and $\phi$ is reminiscent to that in multiferroics \cite{wang}, where electric polarizability (analog of $A^0$) and chiral magnetic order (analog of $\phi$) coexist and mutually interact.

The state proposed here provides the most direct way to realize CSL in a cold atomic setup. This may be done 
in e.g. optical lattices with two cites per unit cell  \cite{greiner1, guinea, tarruell}, which may be tuned into the moat regime. Strongly repelling bosons at half-filling  should form CSL with spontaneosly broken TRS and the 
chiral edge state.     The latter can be  in principle  detected with the recently developed technique of the sudden decoupling~\cite{atala}. 
Accordingly, the density-density correlation-function decays as inverse square distance along the edge, while 
it decays exponentially being measured in the bulk.  Moreover, the momentum distribution function manifests itself in the form of the ``Bose surface'' \cite{Varney-2012}, which may be  revealed in the time of flight experiments. This property, along with real space density asymmetry on sublattices $A$ and $B$ can also be probed in cold atom settings.

Authors thank I. Affleck, L. Balents, E. Demler, V. Galitski, D. Huse, and O. Starykh for illuminating discussions, and Z. Zhu for the providing numerical data prior to their publication.
This work was supported by DOE contract DE-FG02-08ER46482 (AK, LG) and by NSF grant DMR1306734 (TS).



\begin{thebibliography}{99}
\bibitem{footnote} At finite temperatures, there is a transition from AF phase into paramagnetic phase with restored $O(2)$ symmetry. Moreover, one might expect transition from AF phase to CSL while increasing $J_2$ also at a finite-$T$. Such a transition restores $O(2)$ and is most likely of the first order.
\end{thebibliography}
\end{document}